# EDIZ: An Error Diffusion Image Zooming Scheme

Soroush Saryazdi[1], Saman Saryazdi[2], Hossein Nezamabadi-pour[3]

*Department of Electrical Engineering, Shahid Bahonar University of Kerman*
*P.O. Box 76169-133, Kerman, Iran*

[1] soroush.saryazdi@gmail.com

[2] saman.s90@gmail.com

[3] nezam@uk.ac.ir

*Abstract*— **Interpolation based image zooming methods provide a high execution speed and low computational complexity. However, the quality of the zoomed images is unsatisfactory in many cases. The main challenge of super-resolution methods is to create new details to the image. This paper proposes a new algorithm to create new details using a zoom-out-zoom-in strategy. This strategy permits reducing blurring effects by adding the estimated error to the final image. Experimental results for natural images confirm the algorithm's ability to create visually pleasing results.**

## I. Introduction

With the advent of speed, size, and performance of digital computers, image processing techniques have grown dramatically in the last few decades. Although the progress in the camera sensors manufacturing has led to the production of high-quality digital cameras, many applications still require higher quality images, this includes satellite imaging, medical imaging and target detection.

Super-resolution is the process of generating an image with a higher resolution than its source. The source can consist of one or more images or frames[1]. This paper focuses on single frame super-resolution. Single frame super-resolution is also known as zooming, interpolation, and image resizing.

The theory of image super-resolution was first introduced by Tsai et al. in 1984[2]. The objective of super-resolution is to produce a high-resolution image and a higher quality than the reference image, using a single image or several images. A lot of research has been done on the issue of increasing image resolution over the past two decades.

Single frame super-resolution tries to produce a higher-quality image using a single image. However, many researchers believe that the information contained in a single image cannot be enough to increase its resolution, and that it is not possible to reconstruct the high-frequency information by interpolating a low-resolution image.

Various single image super-resolution methods have been proposed so far. They could be classified into three categories: interpolation based method, enhancing based methods and reconstruction based methods.

The first category estimates the missing pixel value using the close relationship between neighboring pixels. In the second category methods, learning algorithms are trained on a database of low-resolution images and their corresponding high-resolution images to estimate the missing details in the low-resolution image.

The last category is methods of reconstruction. These types of methods do not change the size of the image and do not add information, but their main purpose is to enhance images whose resolution was increased by other methods.

Recently, a large number of interpolation approaches have been proposed to improve perceptual quality of reproduced images by considering edge features. Battiato et al. [3] propose a method that takes into account information about discontinuities and sharp luminance variations while doubling the input picture using a nonlinear iterative procedure of the zoomed image with limited computational resources. In [4], an image zooming method based on vector quantization approximation is proposed. The unknown pixel values on the image are interpolated by using a vector quantization codebook based on their local information. Deng et al. [5] use two classes of approximated Heaviside functions (AHFs) to represent smooth and non-smooth components of the input image separately. In [6], the local geometric structures of the two manifolds constructed by low-resolution (LR) and high-resolution (HR) patches are considered to be nonlinear and the original LR and HR patch features are mapped onto the underlying high-dimensional spaces by two nonlinear mappings. Hung and Sieu [7] propose a single-frame super-resolution algorithm using a finite impulse response (FIR) Wiener-filter, where the correlation matrices are estimated using the nonlocal means filter. Saryazdi et al. [8] propose an interpolation based super-resolution algorithm using an adaptive kernel according to the image contents.

Although these methods are efficient for real-time applications, the quality of the reconstructed images is unsatisfactory in many cases. Reconstruction-based methods use priori to solve the problem. They can reconstruct the high frequency texture and suppress false contours, but the results are not satisfactory in a higher magnification. Learning-based methods use machine learning algorithms to solve the image super-resolution problem.

Recently, many learning methods are being proposed to enhance single-image resolution. In most of these methods, a training set is used to learn to produce a high-resolution image given a low-resolution image [9-15].

Interpolation based image zooming methods provide a high execution speed and low computational complexity, but the quality of the zoomed images using these methods is unsatisfactory in many cases. To overcome this drawback, we propose a zoom-out-zoom-in procedure to estimate the missing details in the zoomed image.

The remainder of this paper is organized as follows. We introduce the proposed method and its components in Section II. Experiments and results are presented and analyzed in Section III, and a conclusion is reached in Section IV.

## II. PROPOSED METHOD

Bi-cubic interpolation based image zooming involves the interpolation of unknown pixels by utilizing a cubic kernel. In this paper we opt for bi-cubic image zooming because the image interpolation by using a cubic kernel is fast and computationally inexpensive. However, it suffers from blurring effect. To overcome the blurring effect, our main idea is to predict the missing details in the zoomed image to achieve a higher resolution image. We do this in a zoom-out-zoom-in procedure. Fig. 1 depicts the block diagram of the proposed method.

Here we assume that the zooming factor is $n$. The proposed algorithm possesses five steps. In the first step, the image size is reduced by subsampling by a factor of $n$ (zoom-out). The second step is to zoom the reduced size image by a factor of $n$ using any interpolation method (e.g. cubic) to achieve the reconstructed image. In the third step, the reconstruction error is calculated by subtracting the reconstructed image from the input image. In the next step, the estimated error is achieved by zooming the reconstructed error by a factor of $n$. Finally, the estimated error is added to the zoomed image. The complete procedure of EDIZ is summarized in the pseudo-code given in Table 1.

The zoom-out-zoom-in procedure finds the weakness of the zooming algorithm applied to the given image, i.e. it calculates the details of the given image that the algorithm is not able to reconstruct. Afterwards, it uses this reconstruction error to estimate missing details in the zoomed image. Because of page limitation, here we opt for only bi-cubic and Lanczos zooming, but any zooming algorithm may be used instead.

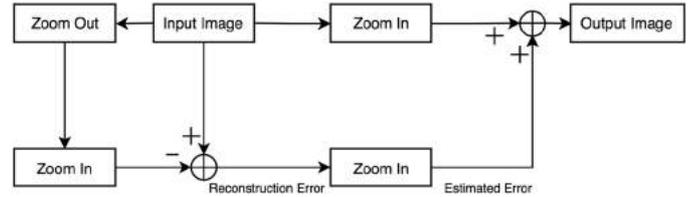

Fig. 1. Block diagram of the proposed method

## III. EXPERIMENTAL RESULTS

To evaluate the performance of the proposed algorithm, we applied it to a wide range of images using bi-cubic interpolation [16]. However, because of page limitation and similar findings among all results, only results of "Lena" and "Tank" are reported here. Fig. 2 shows these results. As these results suggest, the proposed strategy provides a higher visual quality than just using bi-cubic, and more details are visible.

As a second experiment, we applied the proposed algorithm using Lanczos interpolation kernel [17], with $a = 3$ and zooming factor of 4, to "Lena" and "Mandril", 512×512 color images. The results are depicted by Fig. 3. To show the differences more clearly, some regions are zoomed and shown in Fig. 4 and Fig. 5. As these figures confirm, the proposed algorithm achieves results that are richer in detail.

**Table 1.**

| **Algorithm** EDIZ |
|---|
| **Input:** Input image $I_{in}$, Zooming factor $n$ <br> **Output:** Output image $I_{out}$ |
| **Procedure:** <br> 1. Calculate $I_{zout}$ by sub-sampling $I_{in}$ by factor of $n$; <br> 2. Calculate $I_{Rec}$ by zooming $I_{zout}$ by factor of $n$; <br> 3. Calculate the reconstruction error ($e = I_{in} - I_{Rec}$); <br> 4. Calculate estimated error ($E_e$) by zooming reconstruction error $e$ by factor of $n$; <br> 5. Calculate zoomed input image ($I_{in\_zoom}$); <br> 6. $I_{out} = (I_{in\_zoom} + E_e)$; |

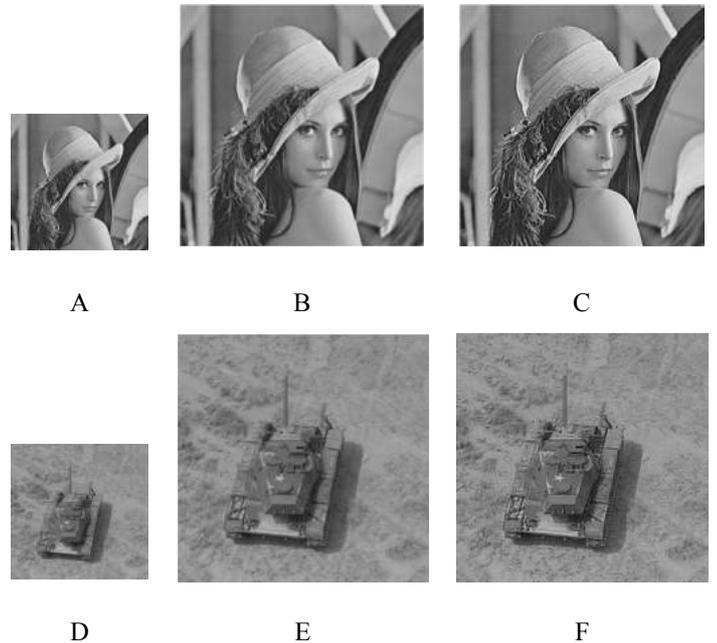

Fig. 2. Comparative results: A, D) input images; B,E) zoomed images by bi-cubic interpolation and a factor of 2; C,F) zoomed image by bi-cubic interpolation and a factor of 2 and using EDIZ.

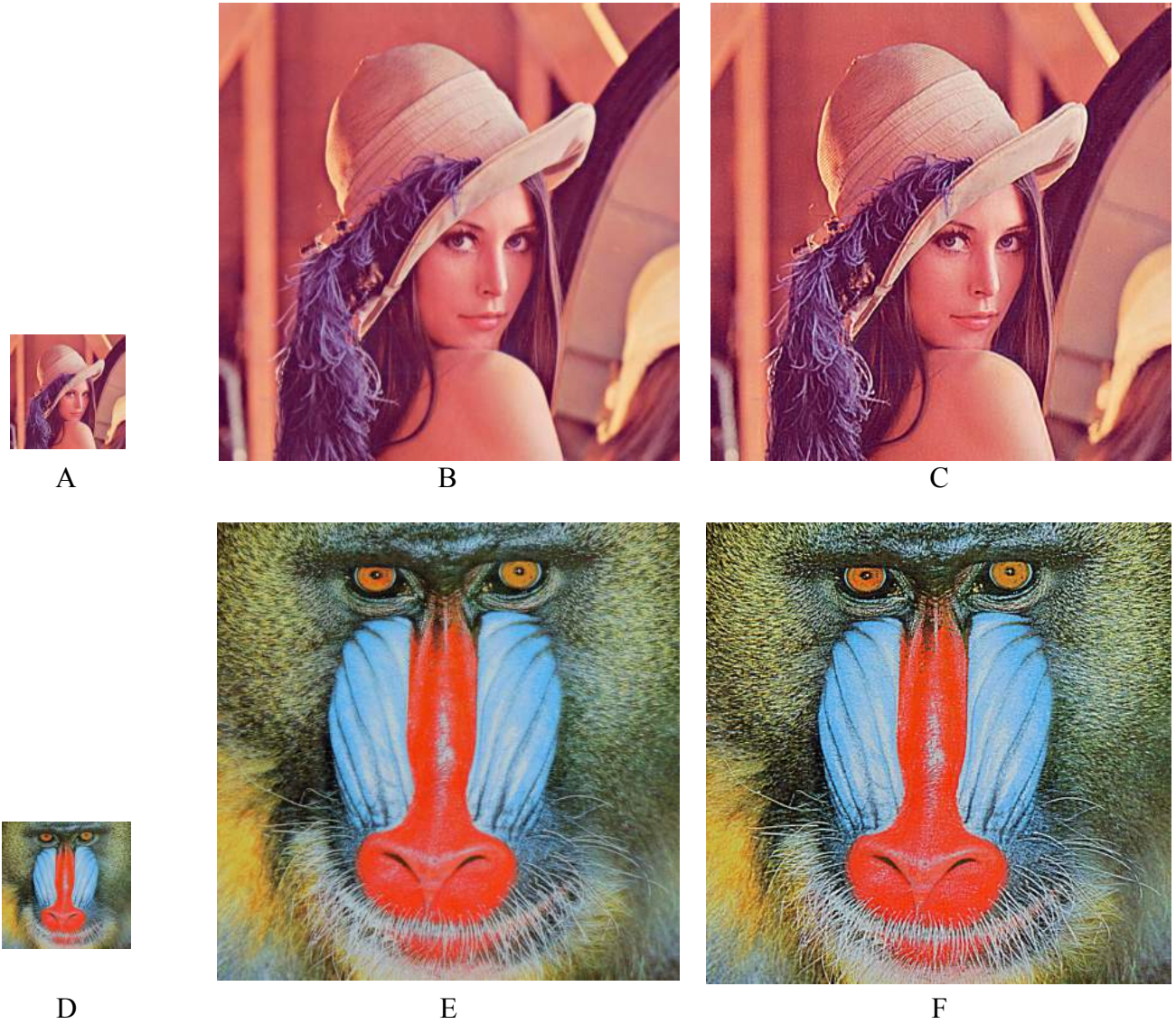

**Fig. 3.** *Comparative results: A, D) input image; B, E) zoomed image by the Lanczos3 interpolation and a factor of 4; C, F) zoomed image by the Lanczos3 interpolation and a factor of 4 and using EDIZ.*

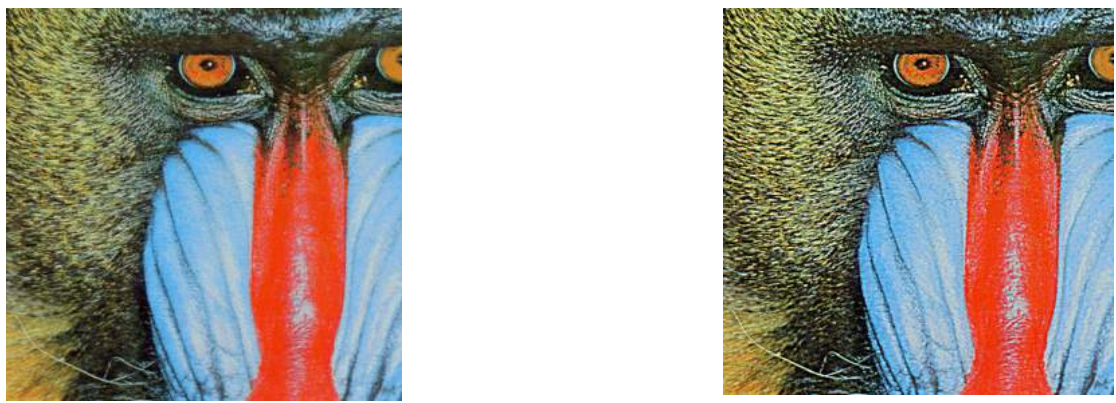

**Fig. 4.** *Details of Mandril in fig. 3: Left) Lanczos3 interpolation; Right) Lanczos3 interpolation and using EDIZ*

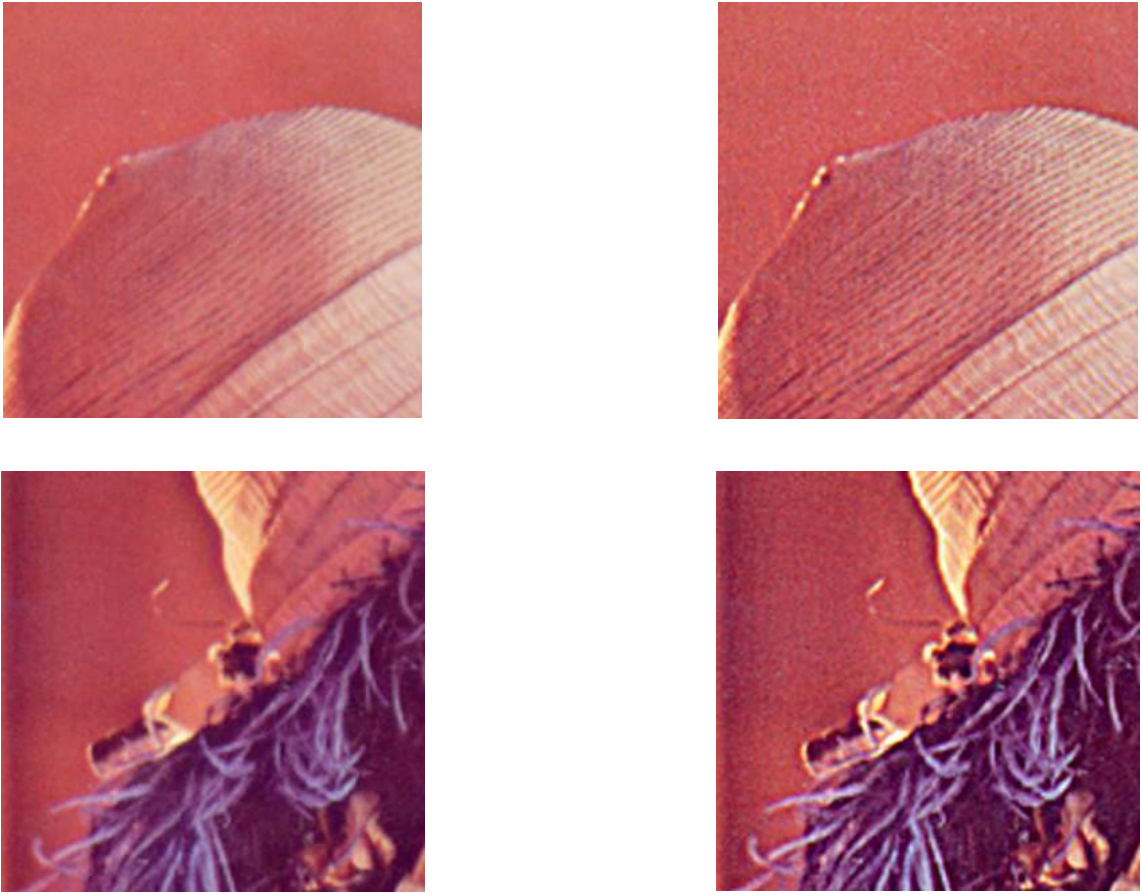

**Fig. 5.** Details of Lena in fig. 3 : Left) Lanczos3 interpolation; Right) Lanczos3 interpolation and using EDIZ

## IV. CONCLUSION

In this paper, we proposed a novel image zooming model named EDIZ to improve interpolation based image zooming quality. The proposed algorithm creates new details using a zooming out-zooming in strategy. This strategy permits reducing blurring effects by adding the estimated error to the final image.

Experimental results for natural images confirmed the good performance of the proposed algorithm and its ability to create visually pleasing results.